\documentstyle[12pt]{article}

\begin{document}

\def\beqra{\begin{eqnarray}} \def\eeqra{\end{eqnarray}}
\def\beqast{\begin{eqnarray*}} \def\eeqast{\end{eqnarray*}}
\def\beq{\begin{equation}}      \def\eeq{\end{equation}}
\def\be{\begin{enumerate}}   \def\ee{\end{enumerate}}
\def\ul{\underline}
\def\indt{\parindent2.5em}
\def\IR{\relax{\rm I\kern-.18em R}}
\def\nd{\noindent}
\def\fnote#1#2{\begingroup\def\thefootnote{#1}\footnote{#2}\addtocounter
{footnote}{-1}\endgroup}

\def\gam{\gamma}
\def\Gam{\Gamma}
\def\la{\lambda}
\def\eps{\epsilon}
\def\La{\Lambda}
\def\si{\sigma}
\def\Si{\Sigma}
\def\al{\alpha}
\def\Tha{\Theta}
\def\tha{\theta}
\def\vphi{\varphi}
\def\del{\delta}
\def\Del{\Delta}
\def\ab{\alpha\beta}
\def\om{\omega}
\def\Om{\Omega}
\def\mn{\mu\nu}
\def\mun{^{\mu}{}_{\nu}}
\def\kap{\kappa}
\def\rsi{\rho\sigma}
\def\beal{\beta\alpha}
\def\til{\tilde}
\def\rta{\rightarrow}
\def\eqv{\equiv}
\def\nab{\nabla}
\def\pa{\partial}
\def\lag{\langle}
\def\rag{\rangle}
\def\caa{{\cal A}}
\def\cb{{\cal B}}
\def\cac{{\cal C}}
\def\cd{{\cal D}}
\def\ce{{\cal E}}
\def\cf{{\cal F}}
\def\cg{{\cal G}}
\def\cah{{\cal H}}
\def\ci{{\cal I}}
\def\cj{{\cal{J}}}
\def\ck{{\cal K}}
\def\cl{{\cal L}}
\def\cm{{\cal M}}
\def\cn{{\cal N}}
\def\cO{{\cal O}}
\def\cp{{\cal P}}
\def\car{{\cal R}}
\def\cs{{\cal S}}
\def\ct{{\cal{T}}}
\def\cu{{\cal{U}}}
\def\cv{{\cal{V}}}
\def\cw{{\cal{W}}}
\def\cx{{\cal{X}}}
\def\cz{{\cal{Z}}}
\def\fo{\hbox{{1}\kern-.25em\hbox{l}}}
\def\rf#1{$^{#1}$}
\def\bx{\Box}
\def\Tr{{\rm Tr}}
\def\tr{{\rm tr}}
\def\BM#1{{\boldmath
\mathchoice{\hbox{$\displaystyle#1$}}
           {\hbox{$\textstyle#1$}}
           {\hbox{$\scriptstyle#1$}}
           {\hbox{$\scriptscriptstyle#1$}}}}
\def\haf{\frac{1}{2}}

\hfill CPP-94-28

\hfill UTTG-14-94

\vspace{1.0cm}

\begin{center}
\large {\bf POINT INTERACTIONS FROM FLUX CONSERVATION}

\vspace{35pt}
\normalsize
Luis J. Boya\fnote{*}{Permanent address:  Departamento de
F\'{i}sica Te\'{o}rica, Facultad de Ciencias, Universidad de Zaragoza, E-50009
Zaragoza, Spain} and E.C.G. Sudarshan

\vspace{.5cm}
{\it Center for Particle Physics,
Department of Physics\\  University of Texas, Austin, TX 78712}

\vspace{.25in}

\begin{minipage}{5.8in}
\abstract{We show that the physical requirement of flux conservation can
substitute for  the usual matching conditions in point interactions.  The study
covers an arbitrary superposition of $\delta$ and $\delta'$ potentials on the
real line and can be easily applied to higher dimensions.  Our procedure can be
seen as a physical interpretation of the deficiency index of some symmetric,
but
not self-adjoint operators.}\end{minipage}
\end{center}

\normalsize
\def\vdown{\vec{\bigtriangledown}}

\vspace{12pt}

\baselineskip=24pt

\parindent2.5em

\nd
(1.)~~~ Point interactions of the delta type have a long history in quantum
physics [1].  In this note we show that the conventional matching conditions
for these potentials can be obtained easily by enforcing the conservation of
the
flux across the discontinuity.

For one-dimensional quantum system with a point interaction at $x =0$, the
continuity equation for the current $\vec{\jmath}$ and the density $\rho$,
namely $\dot\rho +div \vec{\jmath} = 0$ becomes
\beq
   j_- \cong j(x<0)=j_+\eqv j(x >0)
\eeq
in a stationary state; the current is
\beq
\vec{\jmath} =\frac{\hbar}{2im}\; (\psi^* \vdown
 \psi- \psi\vdown \psi^*)\rta \frac{i}{2}\left| \begin{array}{cc}
\psi & \psi^* \\ \psi' & \psi^{\prime*}
\end{array}\right|\;.
\eeq

There are essentially {\it four types} of solutions  to (1) and (2).  If the
flux is zero, we can consider the point $x=0$ as an infinite wall, and we have
two families of total-reflection solutions, labeled by a (constant) phase
shift, namely
\beq
\psi_\al^I(x)=\left\{ \begin{array}{cc} e^{ikx} + e^{i\al}e^{-ikx}& x<0
\\ 0& x> 0 \end{array}\right. ~~~\qquad \psi_\beta^{II}(x)= \left\{
\begin{array}{cc}0 & x<0 \\ e^{-ikx}+e^{i\beta}e^{ikx} & x>0
\end{array}\right.
\eeq
Notice that for generic $\alpha, \beta$, neither $\psi (x)$ nor
$\psi^{\prime} (x)$ vanish at $x=0$, but the flux does.

\nd
(2.)~~~ For non-zero flux, we have another two-parameter family.  Let us
{\it assume} first
\beq
\psi(0-)=\psi(0+)
\eeq
 with perhaps discontinuous $\psi'$ from (1) and (2)
\beq
\psi(0)\;\rm disc\,\,\psi^{*'}(0)- \psi^*(0) \;{\rm disc}\, \psi'(0)
\Rightarrow
\frac{{\rm disc}\,\psi'(0)}{\psi(0)}=\mbox{real const.} = g
\eeq
where disc $f(0) \equiv f(0+) - f(0-)$.

Eq. (5) characterizes $a\; \delta (x)-$ potential of strength $g$.  In fact,
for the scattering situation
\beqra
\psi(x<0) &=& e^{ikx}+b(k)e^{-ikx},\;\;~\psi(x>0)
=(1+f\,(k))e^{ikx}\,,\nonumber
\\
\hat\psi(x<0) &=& (1+\hat f(k))e^{-ikx},\;\;~\hat\psi(x>0)=e^{-ikx}+\hat
b(k)e^{ikx}
\eeqra
we obtain from (4) and (5) the well
known [2] S-matrix
\beq
S(k)\eqv\left(\begin{array}{cc} 1+f(k) & \hat b(k) \\
b(k)&1+\hat f(k)\end{array}\right)  =\left( \begin{array}{cc} 2 ik & g\\ g&
2ik\end{array} \right)\;\frac{1}{2ik-g}\,.
\eeq
The pole at $k=-ig/2$ represents a {\it
bound state} (for $g<0$) or an antibound state (for $g>0$).

\nd
(3.)~~~The {\it fourth} family of solutions is obtained by imposing the
alternative conditions
\beq
{\rm disc} \,\psi(0)=g_1\psi'(0),\qquad {\rm disc}\,\psi'(0)=0\;,
\eeq
in which case the S-matrix becomes
\beq
S(k) = \left(\begin{array}{cc} 2 &-g_1ik\\ -g_1ik&2\end{array}
\right)\;\frac{1}{2-ig_1k}
\eeq
which is the scattering conventionally ascribed to a
$\delta^{\prime}(x)-$ potential [3]; it also supports a single {\it bound}
state (for $g_1 <0$) or antibound state (for $g_1 >0$).

Notice that the $\delta(x)-$ potential is blind to the odd wave, $f(k)=b(k)
\Rightarrow \delta_- (k)=0$, and that the $\delta^{\prime}(x)-$ potential
proceeds exclusively in odd wave, $f(k)= - b(k) \Rightarrow \delta_+ (k)=0$.
Here, $\delta_\pm (k)$ are the even/odd-phase shifts of the one-dimensional
partial waves [4].

\nd
(4.)~~~Our analysis allows logically for a superposition of $\delta (x)-$
and
$\delta^{\prime}(x)-$ potentials which seem to have been so far overlooked in
the literature.  Namely, define $\Phi(x)$ and $\Psi(x)$ by
\beq
\Phi(x) = \cos\al\psi(x)+\frac{1}{m}\sin\al \psi'(x)\;~ \Psi(x)= - m
\sin\al\psi(x)+\cos\al\psi'(x)
\eeq
where $m$ is a quantity with the dimensions of an inverse length.
Then $\Phi$
and $\Psi$ can substitute by $\psi$ and $\psi'$ in (2) provided they are
real since
\beq
\det \left(\begin{array}{cc}\cos\al & +\sin\al/m \\ -m \sin\al &\cos\al
\end{array} \right) =1\,.
\eeq
Now we define the general problem by
\beq
{\rm disc}\,\Phi(0)= 0 \qquad \qquad {\rm disc} \,\Psi(0)=g\Phi(0)
\eeq
and solve for $b, f, \hat{b}$ and $\hat{f}$ of eq. (6); the calculation
is straightforward, yielding
\beq
S(k)=\left(\begin{array}{cc} 2ik & g (\cos \al - \frac{ik}{m} \sin\al) \\
 g(m\cos \al-ik\sin\al)^2 &
2ik\end{array}\right)\;\frac{1}{2ik-g\left(\cos^2\al+\frac{k^2}{m^2}\right)
\sin^2\al}.
\eeq
which interpolates naturally between the
$\delta(x)-{\rm potential},\;\cos\al=1,\;\sin\al=0$  eq. (7); and the,
$\del'(x)-{\rm potential},\;
\cos\al=0,\;\sin\al=1$ eq. (9) with $g=-g_1$.

\nd
(5.)~~~Some features of formula (13) are worth comment.

\begin{enumerate}

\item[{\bf 1.}] $f(k)=\hat{f}(k)$, as demanded by time-reversal invariance[5];
however,
$b(k) \neq \hat{b}(k)$ except in the extreme cases $\delta$ or
$\delta^\prime$.

\item[{\bf 2.}] $\psi_{R=0}(x)=0$ except in the $\delta^\prime (x)$ case, when
$\psi_{R=0}(x)=1$ .

\item[{\bf 3.}] $S$ is, of course, unitary; its spectrum determines the
eigenphase shifts
\beq
\exp2i\del_1=\frac{2ik+g(\cos\al+\frac{k^2}{m^2}\, \sin^2\al)}{2ik-g(\cos^2\al+
\frac{k^2}{m^2}\,\sin^2\al)},\;\,~~\exp 2\del_2=1\,.
\eeq
  This result is worth stressing: {\it our family of
interactions proceeds in a single partial wave, the ``orthogonal'' one is not
affected by the potential}.  This is in consonance  with the simplicity
of the S-matrix, eq. (13): potentials which produce single-mode interaction
have
particularly simple pole structure in the S-matrix[6].  This includes the
delta potential (only even wave), the delta prime (only odd waves), the
``solitonic'' potential
$V(x)=-\ell(\ell+1)\;{\rm sech}^2x,\ell=0,1,2,\ldots$
(only forward scattering) and the one-dimensional
Coulomb potential (only odd-wave interaction).

\item[{\bf 4.}]  For $\sin \alpha \neq 0$ (i.e., excluding the $\delta
(x)\,{\rm
case}$), the two poles of $S$ are given by
\beq
k=im^2\left(
1\pm\sqrt{1+\left(\frac{g^2}{m^2}\right)\cos^2\al\sin^2\al}\right)\;\Big/b\sin^2
\al
\eeq
so there is always a bound state {\it and} an antibound state, for any sign of
$g$, in the mixed case $0\neq\al\neq\pi/2$.  We already remarked that in
the pure cases  $(\al=0$ or $\al=\pi/2)$ there is only one pole, meaning either
a
bound or antibound state.

\item[{\bf 5.}]  The eigenvector of the zero-phase shift is readily seen to be
\beq
V=\left(\begin{array}{c}i\frac{k}{m}\,\sin\al+\cos\al\\ i k
\sin \al -m\cos\al \end{array} \right)
\eeq
and depends only on tan $\al$, say, not on $g$; in particular at
low energies
$V\simeq\left(1 \atop-1\right)$, that is, the {\it odd} wave is not affected,
corresponding to the pure $\delta$ case; at high energies
$V\simeq\left(1 \atop 1\right)$, characteristic of
the $\delta^\prime$ potential, with no force in the even channel.  This is a
sensible result, because the scale dimension of the $\delta (x)$ is 1, but
that of our $\delta^\prime$ is 3 (when dim [momentum] $=+1$).  Note that the
naive dimension of the $\delta^\prime$ would be 2, not 3!

\item[{\bf 6.}]  The reasons to call the matching conditions (8) a
$\delta^\prime (x)-$ potential are obscure; in fact, for the $\delta$ case one
can {\it derive} conditions (4) and (5) by integrating the Schr\"{o}dinger
equation across the discontinuity; this is not so for the $\delta^\prime (x)$.

Also, it is easy to show that a ``regularized'' $\delta^\prime (x)$ potential
\beq
g\lim_{a\rta\tha} \,\frac{1}{a}\,  \{\del (x+a)=\del (x) \}
\eeq
with renormalized  coupling $g$, leads to the conventional
$\delta (x)$ (not $\delta^\prime (x)$!) potential.[7]

The rationale to call conditions (8) a $\delta^\prime (x)$ is that, writing
the Schr\"{o}dinger equation \break
$\psi''+\epsilon\psi=g\delta'(x)\psi,\;\psi''$
is proportional to $\delta^\prime$,
hence $\psi^\prime$ to $\delta$ and $\psi$ to the step function.  Hence,
heuristically,
$\psi^{\prime \prime}$ and $\psi^\prime$ are ``continuous'' at the
singularity, but $\psi$ makes a jump, i.e., conditions (8).
Notice that the
naive $\delta^\prime (x)$ would have dimension $+2$ so it would potentially
be scale invariant, whereas the $\delta'$ we are using has dimension
three; in fact, no trace of scale invariance remains in the $\delta'$ S-matrix,
eq. (9).

\item[{\bf 7.}]  It is not difficult to extend these results to higher
dimensions; we state only the $d=3$ result.[1]  The analogue of eq. (5) is now
\beq
u'/u|_0={\rm const.} \eqv -\frac{1}{a},
\eeq
where $\psi(r)=u(r)/r$ and $u_0(0)=0$; as
\beq
\psi(r) = \frac{1}{r}\,u(r);
\eeq
Since $u=A \sin(r+\delta_0)$, the ``coupling  constant'' determines the phase
shift by
\beq
k\cot \,\del_\tha=- 1/a\,.
\eeq
In this case, $a$ is called the scattering length.  The $d=2$ case has
been the subject of some recent papers[8] and we refer the reader to them.

\item[{\bf 8.}]  The rigorous treatment of the contact potentials entails the
theory of extensions of symmetric,   non-self-adjoint operators, which
started with a paper of Fadeev and Berezin.[9]  But self-adjointness of the
Hamiltonian implies unitarity of the evolution operators, and also of the
S-matrix, which, in turn, is guaranteed by flux conservation; so there is not
much surprise that the families of extensions of the kinetic energy operator
$D=-d^2/dx^2$ acting on $\IR^n-\{0\}$ would coincide with the families of
matching conditions, which we have worked out in detail for the $d=1$ case.[10]

\end{enumerate}

\vspace{12pt}
\noindent
\section*{Acknowledgements}

\parindent2.5em

Luis J. Boya thanks Professor George Sudarshan and the Theory Group of
the University of Texas for their hospitality and partial support.  He is also
grateful to the Spanish CAICYT for a travel grant.  This work was supported by
the Robert A. Welch Foundation and NSF Grant PHY 9009850.

\clearpage

\section*{References}
\begin{enumerate}
\item  S. Albeverio {\it et al.}, ``Solvable Methods in Quantum
Mechanics,'' Springer (Berlin 1988).

\item  See e.g. K. Gottfried, ``Quantum Mechanics,'' Benjamin (New York,
1966), p. 50.

\item  P. Seba, Rep. Math. Phys. {\bf 24}, 111-120 (1986).

\item J. H. Eberly, Am. J. Phys. {\bf 33}, 771-773 (1965).

\item  L. D. Faddeev, Amer. Math. Soc. Translations {\bf 2}, 139-166
(1964).

\item  L. J. Boya, A. Rivero and E.~C.~G. Sudarshan, ``Single Wave
Interactions,'' University of Texas preprint (October 1994).

\item  P. Seba, Ann. Phys. (Leipzig) {\bf 44}, 323-328 (1987).

\item  B. R. Holstein, Am. J. Phys. {\bf 61}, 142-147 (1993).

\item  L. R. Mead and J. Godines, ibid. {\bf 59}, 935-937 (1991). \\
P. Gosdzinsky and R. Tarrach, ibid. {\bf 59}, 70-74 (1991).

\item  F. A. Berezin and L. D. Faddeev, Soviet Math. (Dokladi) {\bf 137},
1011 (1961); Eng. Trasl. {\bf 2}, 372-375 (1961).

\item  M. Carreau, J. Phys. A: Math. Gen. {\bf 26}, 427-432 (1993).
\end{enumerate}
\end{document}